\documentclass[twocolumn]{aa}
\usepackage{graphicx}
\begin{document}
   \title{The Araucaria Project.
          Bright Variable Stars in NGC\,6822 from a Wide-Field Imaging Survey
          \thanks{Tables 2 and 3 are only available in electronic form
at the CDS via anonymous ftp to cdsarc.u-strasbg.fr (130.79.128.5)
or via http://cdsweb.u-strasbg.fr/cgi-bin/qcat?J/A+A/.
Complete figures 1 and 2 are available in the on-line version.}
          }


  \author{R.E.\,Mennickent \inst{1}
 W.\,Gieren \inst{1}
 I.\,Soszy{\'n}ski \inst{1}\fnmsep\inst{2}  \and
 G.\,Pietrzy{\'n}ski \inst{1}\fnmsep\inst{2}
}

 \institute{Universidad de Concepci{\'o}n, Departamento de F\'{\i}sica,
 Casilla 160--C, Concepci{\'o}n, Chile\\
 \email{rmennick@astro-udec.cl, soszynsk@astro-udec.cl, pietrzyn@hubble.cfm.udec.cl, wgieren@astro-udec.cl}
 \and
 Warsaw University Observatory, Al. Ujazdowskie 4,00-478, Warsaw, Poland
 }

\date{Received ; accepted }

\abstract{We have performed a search for
variable stars in the dwarf irregular galaxy NGC\,6822 using
wide-field multi-epoch VI photometry
down to a limiting magnitude $V$ $\sim$ 22. Apart from the Cepheid
variables in this galaxy already reported
in an earlier paper by Pietrzynski et al. (2004),
we have found 1019 "non-periodic" variable stars,
50 periodically variable stars
with periods ranging from 0.12 to 66 days and
146 probably periodic variables.
Twelve of these stars are eclipsing binaries and
fifteen are likely new, low-amplitude Cepheids. Interestingly,
seven of these Cepheid candidates have periods longer than 100 days,
have very low amplitudes (less than  0.2 mag in $I$), and are very red.
They could be young, massive Cepheids still embedded in dusty envelopes.
The other objects
span a huge range in colours and represent
a mixture of different types of luminous variables.
Many of the variables classified
as non-periodic in the present study may turn out to be {\it periodic} variables once a much
longer time baseline will be available to study them.

We provide the catalogue of photometric parameters
and show the atlas of light curves for the new variable stars.
Our present catalogue
is complementary to the one of Baldacci et al. (2005) which has focussed on very short-period
and fainter variables in a subfield in NGC 6822.

   }
   \titlerunning{Bright Variable Stars in NGC\,6822}
   \authorrunning{Mennickent et al}
   \maketitle
%

\section{Introduction}

NGC\,6822 is a relatively well studied dwarf irregular galaxy of the Local Group.
In recent years its content of variable stars has been investigated by several authors, mainly
with emphasis on distance indicators, like Cepheids and RR Lyrae stars,
and short period variables.
Antonello et al. (2002a, 2002b) discovered 130
variable stars in NGC\,6822, 21 of them Cepheids, 18 other periodic
variables and 91
irregular or semiregular variables. Their study was based on a 3-year
monitoring of a small (3.8 x 3.8 arcminute) field of NGC\,6822.
In a series of papers,
Clementini et al. (2003),
Baldacci et al. (2004a, 2004b), and
Baldacci et al. (2005), using deep  VLT images of a field of
6.8 x 6.8 arcminute, reported 390 candidate variables.
They used data taken on 3 half-nights distributed over 5 days
in August 2001. They classified the variables in Main Sequence Variables (MSV, 36 stars),
Classical Instability Strip Variables (CISV, 160 stars) and
Red Giant Branch Variables (RGBV, 66 stars).
They also found 6 eclipsing binaries. As they noted,
their data distribution was optimized for searching periods
in the range of 0.2-0.7 days. On the other hand,
Pietrzy{\'n}ski et al. (2004, hereafter P04)
monitored the whole galaxy in the $V$ and $I$ bands
during 77 nights, spread over nearly one year, with the
aid of mosaic images  (35 x 35 squared arcminutes) taken with the Warsaw 1.3 m telescope
at Las Campanas Observatory, discovering 116 Cepheids
with periods ranging from 1.7 to 124 days. These authors determined,
using the reddening-free Wesenheit index, a distance modulus for NGC\,6822
of 23.34 ($\pm$ 0.04 statistical and $\pm$ 0.05 systematic) mag.
In the present paper, we use the photometric database provided by P04 to
search for variable stars in NGC\,6822. The investigation presented in this paper
complements the previous ones since it is the first simultaneous
monitoring of the whole galaxy during a time interval of almost one year, allowing us
to detect and characterize the population of relatively long-period, bright variables
from these images.

In Section 2, we give a review of the observations and explain our searching tools and selection methods.
In Section 3, we show the results of our investigation, and discuss them
in Section 4. Conclusions  and future prospects are outlined in Section 5.

\section{Observations and selection of variable stars}

The CCD $VI$ images presented in this paper were
obtained with the Warsaw 1.3 m telescope at Las Campanas Observatory.
Most of the observations were obtained in 2002 between July 3 and
November 21. Additional observations were made during three
photometric nights in 2003 May in order to accurately calibrate the data
and improve the periods of the detected variable stars.
The time baseline is about 330 days, total observing nights 77 and the
typical separation between consecutive images is 1 or 2 days.
Details on the observations, and
on the adopted observing
procedures can be found in P04.
Our present search for variable stars was
carried out on the $I$ filter dataset.  All our magnitudes were calibrated
and corrected for the differential photometric zero point across the entire
field of view (P04). Color corrections were
applied using the sum of foreground and intrinsic
reddening for NGC\,300, $E(B-V)$ = 0.36,
derived by McGonegal et al. (1983) and the reddening law of
Schlegel et al.\,(1988), viz.\,
$A_{V}= 3.24\,(B-V)$ and $A_{I}= 1.96\,E(B-V)$.
The absolute magnitude $M_{I}$ was calculated using the
distance modulus for NGC\,6822 derived by P04, viz.\,
$(m-M)_{0} = 23.34$ mag.

The $I$-band light curves of the 
more than one hundred thousand
(107389) photometrically identified
stars in NGC\,6822 were
extracted from our database, yielding robust estimates for
the mean magnitude $<I>$ and standard deviation $\sigma$.
In order to achieve this step, a special algorithm was used which is
less sensitive to outliers  than the usual ones.
The c-code is freely available
at http://www.astro.lu.se/$\sim$stefans/aprog.html. We adopted the following
procedure
independently for each of the 8 chips of the 8k x 8k detector
(see details of the instrumentation in http://ogle.astrouw.edu.pl).
The cloud of pairs $\sigma-<I>$ were passed by a smoothing function
$f(I)$ representing the standard deviation
as a function of magnitude.  The $f$ function turned to be approximately 0.007, 0.008,
0.018, 0.05, 0.10, 0.22 and 0.47 mag. for $I$ = 16, 17, 18, 19, 20, 21 and
22 mag., respectively, almost independent of the chip number.
Stars with $\sigma$ values larger than
$1.1f$ were picked out as variable star candidates. This procedure
left us with 22631 stars, whose light curves were visually inspected in order to
reject false detections due to artifacts of the analysys method.
These artifact might depend on a number of effects, ranging from
hot pixels to cosmic rays or blending of images (coupled with variable seeing).
The light curves were also inspected for periodic variability using
fast and reliable period-searching algorithms based on the algorithm of
analysis of variance (Schwarzenberg-Czerny 1989) and Discrete Fourier
Transform. We calculated the parameter
$S/N$, defined as the ratio between the amplitude of the highest peak in the periodogram and the
noise amplitude, the latter one was calculated as the average of the amplitudes of all peaks in
the periodogram. A safe criterion to establish a significant frequency  seems to be
$S/N > 4$ (Breger et al. 1993, Aerts \& Kolenberg 2005). The figure is also an useful
reference level to evaluate the level of variability detected in the stars,
especially those showing "non-periodic" variability (see next section).

\section{Results}

In addition to the Cepheids already reported by P04, we
found a total of 1019 "non-periodic" variable stars, 146 pvrobably periodic variables
and 50 periodically variable stars. Twelve of them are eclipsing binaries, with periods
between 0.12 and 66 days. "Non-periodic" variables are those without
a clear peak in the periodogram,
but with a periodogram clearly not compatible with noise  (i.e. with $S/N$ $\ga$ 4).
The stars classified as "probably periodic" presented a peak in the periodogram, but it is
not well defined; for these stars, we were able to determine a {\it time scale
of variability}, rather than a true periodicity.
Many of these objects could be periodic, but the time baseline spanned by the present data
is not long enough to
decide this question unambiguously.
On the other hand, periodic stars showed a clear
peak in the periodogram and their periodicity is well established.
Cross-correlating our sample stars with previous surveys in
NGC\,6822 we found three stars in common with
Antonello et al. (2002a),
viz.\, their stars  V\,0130 (1876$-$2),
V\,2881 (765$-$3) and V\,1578 (5874$-$6), and
two
other stars in common with  Baldacci et al. (2005); their
stars V\,112 (710$-$3) and V\,215 (1363$-$3).
For the periodic stars, we determined the ephemeris
for the time of maxima or the main minimum in the case of eclipsing binaries.
The catalogue of variable stars, containing
their mean photometric properties and periods, is given in Tables 1, 2 and 3.
The phased light curves for the periodic variables are
presented in Fig.\,1. The light curves for the probably periodic variables
are displayed in Fig.\,2.

\section{Discussion}

\subsection{The periodic and probably-periodic variable stars}

Looking at the colour-magnitude diagram of the periodic and probably periodic
variables (Fig.\,3), we observe
that they span a huge range of colours, from $V-I$ = -0.4 to 2.7.
They are also quite luminous, with absolute visual magnitudes between -8.6
(the case of 32$-$5, the brightest star in our sample) and -2.6 (the star 5574$-$1).
The stars seem to be placed in well-defined regions of the CM diagram.
For instance, we distinguish five "blue" variable stars, which are inside the upper left
rectangle in Fig.\,3. These stars have relatively blue $V-I$ colours, 3 of them
are placed close to the region of the O-type main sequence stars.
In addition, 8 stars are placed above the yellow supergiant track,
and 76 stars are red variables located in the asymptotic giant branch region
indicated by the rectangle at the right side of the Fig.\,3.
Most of these later stars are very slow variables, with time scales of variability between
71 and 188 days, although the upper limit is severely biased by our time baseline.
In addition, we have found a large percentage of our variables mostly
occupying the upper part of the classical Cepheid instability strip, and a small
fraction of stars near the late-K supergiant track. It is interesting to note that
only
about 1/3 of the stars located in the classical Cepheid instability strip
were found to be Cepheids, as indicated by their typical light curve shapes and
colours. Contamination of the sample by foreground stars is not large enough to
account for this effect (see Section 5), so probably there are other kinds of
variable stars apart from Cepheids inside their pulsation instability zone
(see also Fig.\,7, and corresponding discussion in Section 4.2,
for additional low amplitude variables in this region).

\subsubsection{Blue variables}

The six stars located in the "blue" part of the CM diagram correspond
to three double eclipsing systems (488$-$3, 803$-$3 and 2516$-$1,
with periods 6.66, 66 and 3.44 days,
respectively) and three periodic stars with periods 0.18 days (377$-$6),
2.67 days (4907$-$6), and 2.61 days (5698$-$6). These later stars show regular
"single-wave" light curves with $I$-band amplitude 0.3$-$0.5 mag.
The periods could be rotational, except for 377$-$6, which could be a short-period
RR Lyrae star of the galactic halo of type RRc.

\subsubsection{New Cepheids}

We looked for new Cepheids in our sample by constructing $I$ versus $\log P$
diagrams for our periodic stars, and overplotting the Cepheids
found by P04. The stars lying close to these Cepheids (9 stars),
were selected as new Cepheid candidates (Fig.\,4). After examining their light curves for
classification consistency (shape, period and amplitude),
we concluded that they are indeed likely to be Cepheids, except
2516$-$1  which is an eclipsing binary.
The new Cepheid candidates  are
labeled   16981$-$2, 2536$-$3, 390$-$6, 1433$-$6, 10900$-$6, 6115$-$7, 6123$-$7 and  6867$-$7.
We did the same exercise with the quasi-periodic variables with periods in excess
of 100 days. We found eight
stars lying close to the $PL$ relationship (Fig.\,4); they are
the variables
210$-$1,  57$-$2, 65$-$3, 6193$-$5,   4641$-$6,   4741$-$6, 4898$-$6 and 449$-$8.
Their long time scale of
variability (larger than 110 days) and very small amplitudes (less than 0.2 mag),
apart from their position in the CM diagram ($V-I$ usually larger than 1.6),
pointed to a classification different from Cepheids.
However, when plotting these objects on the period-luminosity relationship
they turn out (except 449$-$8, which is clearly an outlier, and has a low luminosity)
to be
almost aligned with the "bona-fide" Cepheids (Fig.\,5).
In the $I$-band $PL$ relation, these 7 stars almost follow the $PL$ sequence
defined by the shorter period stars, but are somewhat underluminous (as compared
to the expected $I$ magnitudes when extrapolating the shorter-period sequence).
In the $V$-band $PL$ relation ($V$ vs $\log P$), the effect is much stronger;
the 7 stars are at least 1 mag fainter than expected if they were
normal Cepheids, assuming that the Cepheid PL relation keeps being linear (in the magnitude vs. logP plot)
up to such high periods. However, in the reddening-free Wesenheit magnitude these 7 objects
lie exactly on the extension of the $PL$ relation defined by all the
normal, shorter-period Cepheids (Fig.\,5).
In the color-magnitude diagram, the seven stars all seem to be too red to lie within the Cepheid
instability strip (Fig.\,3). However, we caution here that the red edge of the
instability strip is currently not well defined; models have still trouble to
incorporate convection correctly into the models. Therefore, it seems not completely out of the
question that these objects are very red Cepheids, very close to the
red edge of the instability strip, and perhaps pushed slightly beyond it by excessive differential reddening.
This would be consistent with their very
low amplitudes. In the period-color diagram ($(V-I)_{0}$ vs $\log P$),
the 7 stars are by
0.5-1.0 mag redder than what would be expected from the prolongation
of the period-color relation defined by the shorter-period Cepheids.
This might argue, in principle,  against a classification as Cepheids; however, the intrinsic dispersion of
Cepheid period-colour relations is large.

As a conclusion at this point, things are a bit confusing; the nature of
these stars is not really clear. While they could be Galactic foreground stars
(but 7 being so similar in their characteristics is hard to imagine), or variable red
supergiants, they could also be very long-period Cepheids. There is very
little known empirically and theoretically about such stars, and their properties.
In the literature, it has been claimed that Cepheids with P $>$ 100 days have a
tendency of departing from the $PL$ sequence of the shorter-period Cepheids,
towards lower luminosities, exactly what we see.
On the other hand, the Cepheid classification is consistent with the low
metallicity of NGC\,6822 (e.g. Clementini et al. 2003) and the
theoretical
expectation that the longest period Cepheids should be found in environments
of low metallicity (Aikawa \& Antonello 2000).
If these seven objects are
very long-period Cepheids, they must be very massive, and therefore, very young
stars; they could then perhaps still be dust-enshrouded to some extent,
which could partly explain why these stars are so red.
It will be very interesting and important
to follow up on these variables because they might give us a clue to find out if such
extremely long-period Cepheids indeed exist, and if so, if they follow the $PL$ relation
defined by the shorter-period (less than 100 days) Cepheids. This is a very important
issue for the use of Cepheids as distance indicators, where the longest-period and
brightest Cepheids are potentially the most useful ones. Little is known about such
possible Cepheids of extremely long periods because of their extreme paucity, and
NGC 6822 could be a key galaxy to study them, if the Cepheid nature of these seven enigmatic objects
is confirmed
by future observations.

\subsubsection{Stars in the yellow supergiant region}

Eight stars are located above the yellow supergiant track in Fig.\,3.
We investigated if these stars
could be related to the yellow hypergiant stars reviewed by de Jagger (1998). Yellow
hypergiants are possibly  evolved stars evolving from the red supergiant phase to the blue phase.
The candidate stars are 1$-$3, 28$-$4, 4584$-$4, 32$-$5, 58$-$5, 6066$-$5, 6140$-$5 and 5668$-$7.
By examination of the light curves of these objects, we find that
6140$-$5 is a near-contact binary with orbital
period 0.34 days. For this period, the components of a contact binary should be much
less luminous, so we conclude that this is a foreground object that does not
belong to NGC\,6822. The other objects have periods 0.4 days  (1$-$3, $A$ = 0.26),
2.8 days (28$-$4, $A$ = 0.06), 5.6 days (4584$-$4, $A$ = 0.08$-$0.05), 43 days (32$-$5,
$A$ = 0.05), 2.8 days (58$-$5, $A$ = 0.10), 8.3 days (6066$-$5, $A$ = 0.31) and
0.19 days (5668$-$7, $A$ = 0.52). As we can see, the periods are rather short,
much shorter than those observed in pulsating yellow hypergiants (hundreds of days).
On the other hand, the periods are also too short to
allow a companion star around a supergiant, so we conclude that
these stars are very likely members of our galaxy. Stars 1$-$3 and 5668$-$7 could
be RR Lyrae stars in the halo of our galaxy.

\subsubsection{The long period variables}

The last group in our discussion corresponds to the red variables
in the right rectangle in Fig.\,3. These stars have 1.3 $<$ $(V-I)_{0}$ $<$
2.8 and -4.1 $<$ $M_{V}$ $<$ -2.0. They are slow variables, with time scales
of variability between 70 and 190 days.
The upper limit is due to our selection mechanism; the application of the
automatic period-searching algorithm hardly finds stars with longer periodicities
due to the time baseline defined by our data. Interestingly, some of these stars
show also variability on shorter time scales (see Fig.\,2).
Their $I$ amplitudes show a smooth distribution
peaking at 0.5 mag (Fig.\,6). These stars are likely
Mira stars, or semiregular variables (Kholopov et al. 1985).
There is no correlation between time scale of variability and colour, amplitude or magnitude
for these stars. Our stars occupy a region of higher luminosity than
the tip of the red giant branch (TRGB) and we do not see
$PL$ relationships for them as those
observed in the Magellanic Clouds for TRGB stars (Fraser et al. 2005).
However, we note that bright stars do not show large amplitude
variability (Fig.\,6). This could be explained if bright stars showing
large amplitudes also show predominantly longer periods, and therefore would escape from detection
in our present study.

\subsection{The non-periodic variable stars }
 
Figure 7 shows the CM diagram for the 1019 non-periodic variable stars.
We observe some stars close to the main sequence, in the region of high-mass stars.
We also find many high luminosity stars in the region of the supergiants of types G0-M2.
$I$-band amplitudes for these stars vary from  0.02 to 1.18 mag, with a mean
of 0.15 ($\pm$ 0.14 std) mag, and no correlation with $(V-I)_{0}$ colour.
The most notable feature of the diagram is the
strong concentration in the region of the red giants. For these red
stars we show the histogram
of the $I$-amplitudes in Fig.\,8, along with their  distribution with
absolute $I$ magnitude. The difference with Fig.\,6 is notable. We observe smaller  amplitudes and no correlation with
luminosity, which confirms our suggestion that the deficit of stars with large amplitudes and
high luminosities observed in Fig.\,6 is an artifact.
Another point of interest is the large number of stars
in the region occupied by the classical Cepheids discovered by P04. It is hard to imagine that all
these stars are foreground galactic stars (see also discussion in Section 5),
so it is probable,
as we already noted in Section 4.2,  that
other kinds of variables are located in the Cepheid instability strip.
Spectroscopy is  needed to establish the nature of these low-amplitude objects.

 \begin{figure*}
   \centering
   \includegraphics[width=15cm]{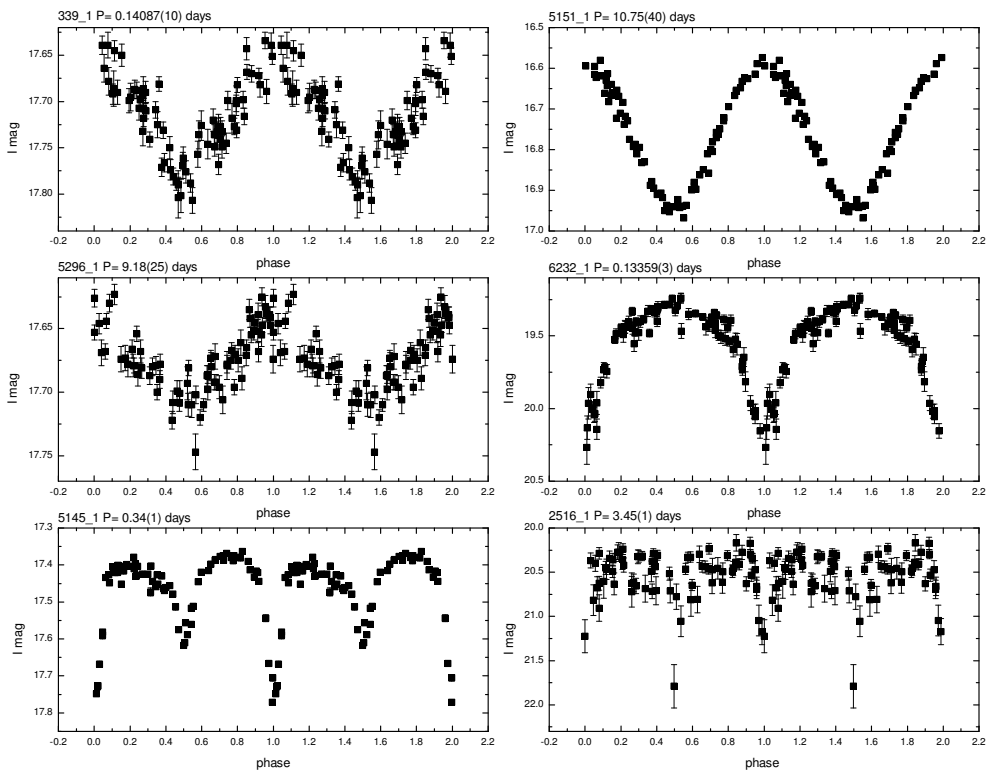}
      \caption{Examples of light curves for periodic variables. The
      full atlas is available in the
      electronic version.
              }
         \label{Fig}
   \end{figure*}

 \begin{figure*}
   \centering
   \includegraphics[width=15cm]{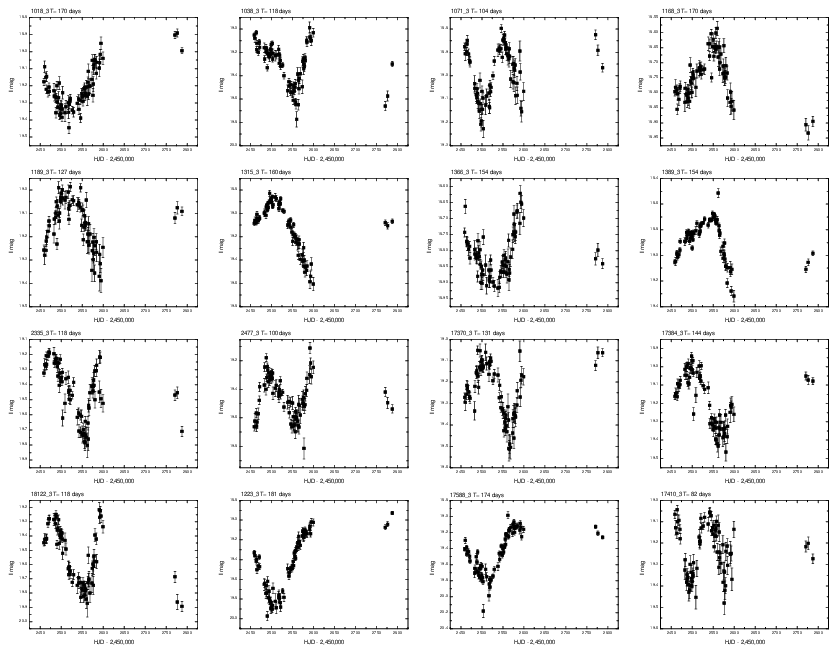}
      \caption{Examples of light curves for probably periodic variables. The
      full atlas is available in the electronic version.
              }
         \label{Fig}
   \end{figure*}

\tiny
 
\begin{table*}
\caption{Periodically variable stars towards NGC\,6822. We give the
amplitude in the I-band ($\Delta I$), period and absolute magnitude, dereddened colours, and $I$ and $V$
magnitudes.
The error in the last digit of the period is given
in parenthesis. An "$e$" accompanying the period means the stars
is an eclipsing binary.  Times for maxima
(or main minima, in the case of eclipsing binaries) are also given.
$e(HJD$) means error in $HJD$ and $MA$ means "multiple aliases".}             
\label{table:1}      
\centering                          
\begin{tabular}{c c c c c c c c c c c}        
\hline\hline                 
$ID$ &alpha(2000) & dec(2000) & $\Delta I$ (mag) & $P$ (days) &$M_{I}$ &$(V-I)_{0}$ &$I$ &$V$ &$HJD$ &$e(HJD)$ \\    
\hline                        
291$-$6  &19:43:55.53 &-14:43:28.9 & 0.70 &   0.29006(9)e        & -5.87&   0.80 &  18.14 &  19.40   & 2459.0477&   0.0014   \\
32$-$5   &19:43:58.13 &-14:33:11.8 & 0.05 &   43 (5)             & -9.32&   0.71 &  14.69 &  15.86  & 2474.4549&   1.7200   \\
234$-$5  &19:43:58.92 &-14:36:38.1 & 0.17 &   5.12 (6) or 2.48 (3)  & -6.10&   0.84 &  17.90 &  19.20   & 2459.1901&   0.1024   \\
241$-$5  &19:44:00.77 &-14:34:33.0 & 0.69 &   0.2546(2)e         & -6.31&   0.76 &  17.69 &  18.91  & 2459.3815&   0.0013   \\
684$-$7  &19:44:02.19 &-14:49:56.4 & 0.36 &   0.11935(10)MA      & -5.12&   0.65 &  18.88 &  19.99  & 2458.9319&   0.0036   \\
382$-$7  &19:44:02.92 &-14:52:00.6 & 0.20 &   0.2760(5)MA        & -6.03&   0.60 &  17.97 &  19.03  & 2459.0798&   0.0072   \\
58$-$5   &19:44:07.15 &-14:31:57.4 & 0.10 &   2.76 (3)           & -8.94&   0.57 &  15.07 &  16.10   & 2459.5427&   0.1380   \\
112$-$7  &19:44:11.63 &-14:55:21.9 & 0.09 &   4.51 (3)           & -7.78&   1.19 &  16.22 &  17.87  & 2462.9592&   0.1353   \\
377$-$6  &19:44:11.76 &-14:44:12.5 & 0.31 &   0.17858(8)         & -6.19&   0.08 &  17.82 &  18.36  & 2458.9957&   0.0018   \\
1433$-$6 &19:44:12.78 &-14:41:08.3 &1.03 &   5.73(6)            & -4.11&   0.48  &  17.77 &  19.15  & 2460.6911&   0.0860   \\
390$-$6  &19:44:13.72 &-14:39:47.1 & 0.13 &50(-3,+10) or 1.017(5)& -6.23&   0.92 &  17.83 &  19.34  & 2489.9253&   0.7500   \\
6220$-$5 &19:44:20.84 &-14:31:45.4 & 0.14 &    0.11985 (5) MA    & -6.18&   1.05 &  16.21 &  17.42  & 2458.9402&   0.0048   \\
6140$-$5 &19:44:26.22 &-14:36:24.4 & 0.60 &   0.34216(5)e        & -7.80&   0.75 &  18.52 &  20.08  & 2458.8904&   0.0021   \\
4983$-$6 &19:44:27.80 &-14:43:00.0 & 0.31 & 0.1470 (1) or 0.0735 (1)  & -5.49&   1.10 &  17.61 &  18.78  & 2458.9420&   0.0029   \\
4786$-$6 &19:44:30.88 &-14:40:00.1 & 0.11 &   3.745(15) MA       & -6.40&   0.71 &  15.99 &  16.80   & 2461.7250&   0.1498   \\
6066$-$5 &19:44:30.88 &-14:38:27.4 & 0.31 &    8.34 (15)         & -8.02&   0.35 &  18.37 &  18.50   & 2462.9778&   0.1668   \\
4794$-$6 &19:44:32.83 &-14:46:01.1 & 0.72 &   0.4972(6)e          & -6.57&   0.40 &  16.71 &  18.59  & 2459.1496&   0.0050  \\
4907$-$6 &19:44:34.24 &-14:42:20.5 & 0.49 &    2.67 (3)          & -5.64&   -0.33&  20.23 &  21.30   & 2461.1760&   0.0267   \\
5663$-$7 &19:44:37.32 &-14:56:03.0 & 0.10 &   3.89 (4)           & -7.30&   1.42 &  21.29 &  -      & 2460.5838&   0.0778   \\
10900$-$6&19:44:40.47 &-14:47:03.9 & 1.00 &    5.23 (10)         & -3.77&   0.61 &  16.04 &  17.23  & 2459.7992&   0.4184   \\
12194$-$7&19:44:40.48 &-14:51:05.3 & 1.96 &    15.95 (40)        & -2.71&   --   &  18.85 &  20.73  & 2465.1790&   0.3190   \\
5668$-$7 &19:44:41.63 &-14:56:35.6 & 0.52 &   0.19369 (10)       & -7.97&   0.73 &  18.88 &  19.00     & 2458.8996&   0.0029   \\
6115$-$7 &19:44:47.15 &-14:49:22.6 & 0.58 &   19 (1)             & -5.15&   1.42 &  20.50 &  21.16  & 2459.0513&   0.3800   \\
5698$-$6 &19:44:48.86 &-14:44:00.3 & 0.30 &   2.609 (16)         & -5.13&   -0.34&  19.79 &  20.87  & 2460.1238&   0.0522   \\
12770$-$6&19:44:49.13 &-14:43:59.1 & 1.64 &    25.35 (1.0)e       & -3.51&   0.20 &  17.18 &  19.76  & 2473.5554&   1.2675   \\
8846$-$7 &19:44:51.95 &-14:54:31.1 & 1.03 &   33 (2)             & -4.21&   0.62 &  18.78 &  20.38  & 2472.6413&   0.9900   \\
6366$-$5 &19:44:54.78 &-14:38:19.2 & 0.17 &    0.13087 (5)       & -6.82&   2.12 &  19.43 &  -      & 2458.8946&   0.0023   \\
6123$-$7 &19:44:55.03 &-14:50:18.8 & 0.30 &   11.5 (3)           & -5.22&   1.14 &  15.73 &  16.53  & 2460.6128&   0.2300   \\
6867$-$7 &19:44:55.57 &-14:50:06.6 & 0.56 &8.49(11) or 0.5296(5) & -4.58&   --   &  18.93 &  19.40   & 2464.7121&   0.5094   \\
1$-$3    &19:44:57.51 &-14:45:02.8 & 0.26 &   0.39669(20)        & -8.27&   0.34 &  20.10 &  21.04  & 2459.1251&   0.0079   \\
803$-$3  &19:45:01.26 &-14:45:40.0 & 0.59 &    66 (3 )e          & -5.07&   0.01 &  17.31 &  19.58  & 2498.7443&   1.2000   \\
2536$-$3 &19:45:04.18 &-14:45:03.9 & 1.20 &   5.10 (5)           & -3.91&   0.48 &  15.84 &  17.58  & 2459.2154&   0.1020   \\
203$-$3  &19:45:04.40 &-14:42:48.6 & 0.10 & 2.83 (1) or 1.539(1)  & -6.69&   1.81 &  18.15 &  19.92  & 2462.2903&   0.0840   \\
77$-$4   &19:45:06.75 &-14:37:07.2 & 0.03 &   21.3 (1.3)         & -8.17&   1.28 &  16.64 &  18.40   & 2460.5098&   0.4260   \\
2516$-$1 &19:45:10.89 & -15:01:58.1&1.62 &   3.45(1)e            &-3.47 &  -0.38  &  18.39 &  18.44  & 2459.1073&   0.0690  \\
339$-$1  &19:45:12.68 &-15:03:05.5 & 0.18 &   0.14087(10)        & -5.86&   1.31 &  15.42 &  16.64  & 2458.9011&   0.0014   \\
197$-$4  &19:45:15.61 &-14:34:59.7 & 0.13 &   1.556(9)           & -7.37&   1.30 &  17.18 &  18.57  & 2461.9783&   0.3110   \\
488$-$3  &19:45:16.36 &-14:45:15.9 & 0.56 &   6.66(3)e           & -5.61&   -0.41&  17.34 &  19.09  & 2462.9829&   0.0330   \\
28$-$4   &19:45:16.55 &-14:37:53.8 & 0.06 &   0.7363(5)          & -8.59&   0.76 &  17.00 &  18.30   & 2461.7467&   0.0221   \\
244$-$2  &19:45:18.29 &-14:50:19.9 & 0.65 &    0.2164 (1)e       & -6.83&   0.93 &  16.76 &  18.41 & 2459.0444&   0.0022   \\
258$-$2  &19:45:21.01 &-14:51:54.5 & 0.16 &    1.84 (1)          & -6.67&   1.29 &  17.67 &  19.55  & 2458.8935&   0.0184   \\
225$-$4  &19:45:25.35 &-14:33:26.7 & 0.69 &   0.40204(20)e       & -7.00&   0.84 &  19.57 &  21.04  & 2459.1897&   0.0040   \\
5145$-$1 &19:45:36.77 &-14:58:00.5 &   0.41&    0.34(1)          &-6.54&    1.28 &  16.56 &  18.08  & 2459.1386&   0.0024  \\
5151$-$1 &19:45:38.02 &-14:59:50.1 & 0.39 &   10.75(40)          & -7.24&   1.18 &  18.77 &  19.48  & 2463.9651&   0.1075   \\
5296$-$1 &19:45:40.97 &-15:00:03.1 & 0.12 &   9.18(25)           & -6.33&   1.42 &  17.13 &  18.46  & 2466.0132&   0.0918   \\
6232$-$1 &19:45:56.02 &-15:00:21.5 & 1.03 &   0.13359(3)e        & -4.44&   1.01 &  15.01 &  16.33  & 2458.9510&   0.0013   \\
16881$-$3&19:45:57.08 &-14:44:13.3 & 0.25 &   0.16566(5)e        & -7.45&   1.06 &  19.90 &  20.84 & 2460.8813&   0.0033   \\
16981$-$2&19:45:58.85 &-14:51:49.2 & 0.32 &    9.18 (20)         & -5.23&   0.25 &  17.44 &  18.30 & 2466.5456&   0.2754   \\
17011$-$3&19:45:59.21 &-14:43:08.8 & 0.17 &    15.7 (6)          & -6.88&   0.87 &  17.47 &  19.21 & 2460.8582&   0.1260   \\
4584$-$4 &19:46:03.38 &-14:31:46.2 & 0.08 &    5.59 (3)          &  -8.9&  0.86 &  20.53 &  20.62 & 2463.4753&   0.0838   \\
 \hline
\end{tabular}
\end{table*}
\normalsize

\begin{table*}
\caption{Candidates for periodically variable stars towards NGC\,6822. Nomenclature is as in
Table 1. This is an excerpt, the whole table is available in electronic form.}             
\label{table:2}      
\centering                          
\begin{tabular}{c c c c c c c c c}        
\hline\hline                 
$ID$ &alpha(2000) & dec(2000) & $T$ (days)& $\Delta I$ (mag) &$M_{I}$ &$(V-I)_{0}$&$I$ &$V$  \\    
\hline
449$-$8  &19:43:56.76 &-15:04:10.5 &178 & 1.34 &    -4.52  & 1.67   &19.48 & 21.62   \\
1135$-$8 &19:43:58.54 &-14:59:25.8 &147 & 1.09 &    -4.08  & --     &19.92 &  -      \\
662$-$8  &19:44:05.08 &-14:57:37.4 &94  & 0.51 &    -4.90  & 2.07   &19.10 &  21.63  \\
664$-$8  &19:44:05.28 &-15:02:43.0 &120 &0.36       &  -5.47 &  1.67&18.54 &  20.67   \\
919$-$5  &19:44:09.79 &-14:36:26.9 &170 & 0.64 &    -4.46  & 1.89   &19.54 &  21.89  \\
757$-$6  &19:44:15.51 &-14:44:10.4 &120 & 0.53 &    -4.77  & 2.12   &19.24 &  21.82  \\
828$-$7  &19:44:15.96 & -14:50:41.2& 182 &0.65   & -4.77   &   2.08 &19.24 &  21.78  \\
482$-$7  &19:44:19.43 &-14:51:31.7 &109 & 0.34 &    -5.05  & 2.09   &18.95 &  21.50  \\
5713$-$8 &19:44:23.14 &-15:02:27.5 &144 & 0.37 &    -4.74  & 2.09   &19.27 &  21.82  \\
6246$-$7 &19:44:28.59 &-14:50:44.6 &141 & 0.72 &    -4.65  & 1.97   &19.36 &  21.79  \\
 \hline
\end{tabular}
\end{table*}

 \begin{table*}
\caption{"Non-periodic" variable stars towards NGC\,6822. This is an excerpt, the whole
table is available in electronic form.
The identification number of the star, the field number, coordinates,
$I$-band amplitude, absolute $I$ magnitude and dereddened
$(V-I)_{0}$ colour are given, as well as the $I$ and $V$ magnitudes.
$S/N$ is the ratio between the amplitude
of the highest peak in the periodogram and the noise amplitude. {\it Note}: Many of these
variables may turn out to be periodic once the time baseline of the observations will
be extended.}             
\label{table}      
\centering                          
\begin{tabular}{c c c c c c c c c c}        
\hline\hline                 
$ID number$ & Field &alpha(2000) & dec(2000) & $\Delta I$  (mag) &$M_{I}$ &$(V-I)_{0}$ &$I$ &$V$ &$S/N$ \\    
\hline
17 & 1 & 19:45:13.32 & -14:58:44.6 & 0.03 & -8.29 & 1.51   &   15.71 &  17.69 &  8.03             \\
554 & 1 & 19:45:07.45 & -14:59:23.6 & 0.09 & -5.06 & 2.42  &   18.94 &  21.83 &  6.56              \\
564 & 1 & 19:45:08.15 & -15:02:16.1 & 0.05 & -4.66 & 1.94  &   19.34 &  21.75 &  4.45              \\
572 & 1 & 19:45:08.99 & -14:57:46.1 & 0.11 & -4.76 & --    &   19.25 &  -     &  6.09             \\
628 & 1 & 19:45:14.76 & -14:58:23.8 & 0.11 & -4.81 & 2.47  &   19.20 &  22.13 &  5.79             \\
660 & 1 & 19:45:18.22 & -14:59:52.2 & 0.09 & -4.75 & 2.28  &   19.26 &  22.00 &  5.24             \\
679 & 1 & 19:45:20.45 & -14:59:38.0 & 0.18 & -5.01 & --    &   18.99 &  -     &  8.94             \\
744 & 1 & 19:45:26.92 & -14:57:41.6 & 0.13 & -4.87 & 2.45  &   19.14 &  22.05 &  7.41              \\
798 & 1 & 19:45:32.34 & -15:01:01.6 & 0.05 & -4.82 & 2.01  &   19.19 &  21.66 &  4.79              \\
892 & 1 & 19:45:00.18 & -15:00:55.2 & 0.53 & -3.64 & --    &   20.37 &  -     &  8.34              \\
\hline
\end{tabular}
\end{table*}

   \begin{figure*}
   \centering
   \includegraphics[width=15cm]{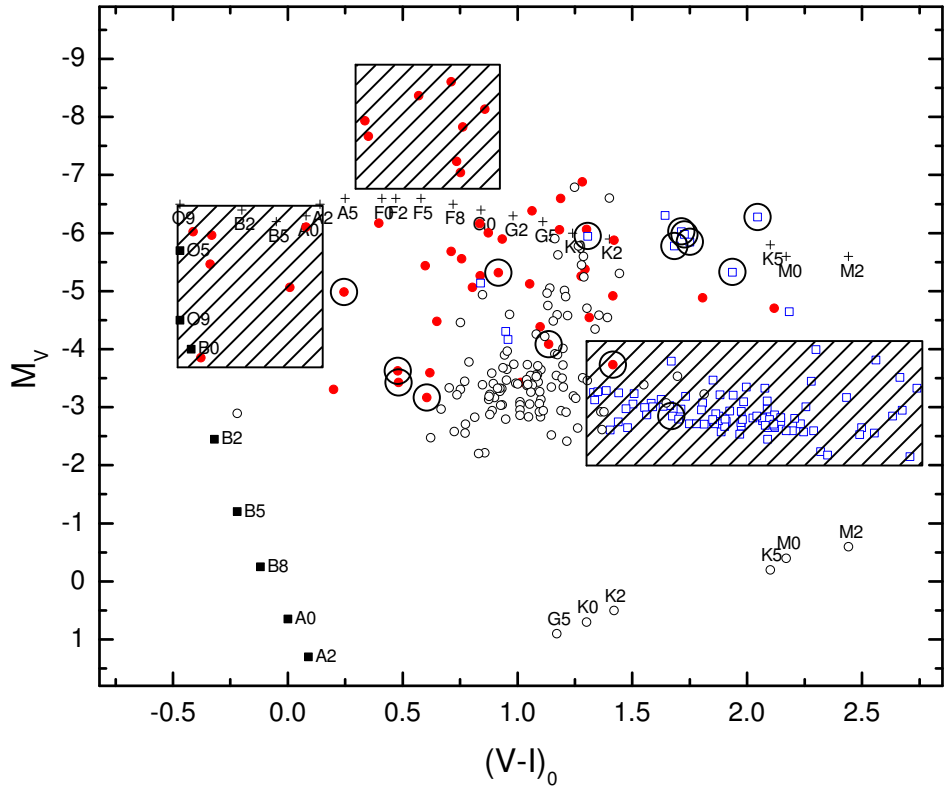}
      \caption{The colour-magnitude diagram for the variable stars towards
NGC\,6822. Small filled circles correspond to periodically variable stars, and open
squares to candidates for periodic variables. Cepheid candidates are indicated by large circles.
The Cepheids reported by
P04 are shown as small open circles.
The position of the upper blue main sequence is
indicated, along with the tracks for giants and supergiants accordingly to
Cox (2000).
The dashed rectangular boxes include the blue, yellow and red stars
discussed in the text.
              }
         \label{Fig}
   \end{figure*}

      \begin{figure*}
   \centering
   \includegraphics[width=15cm]{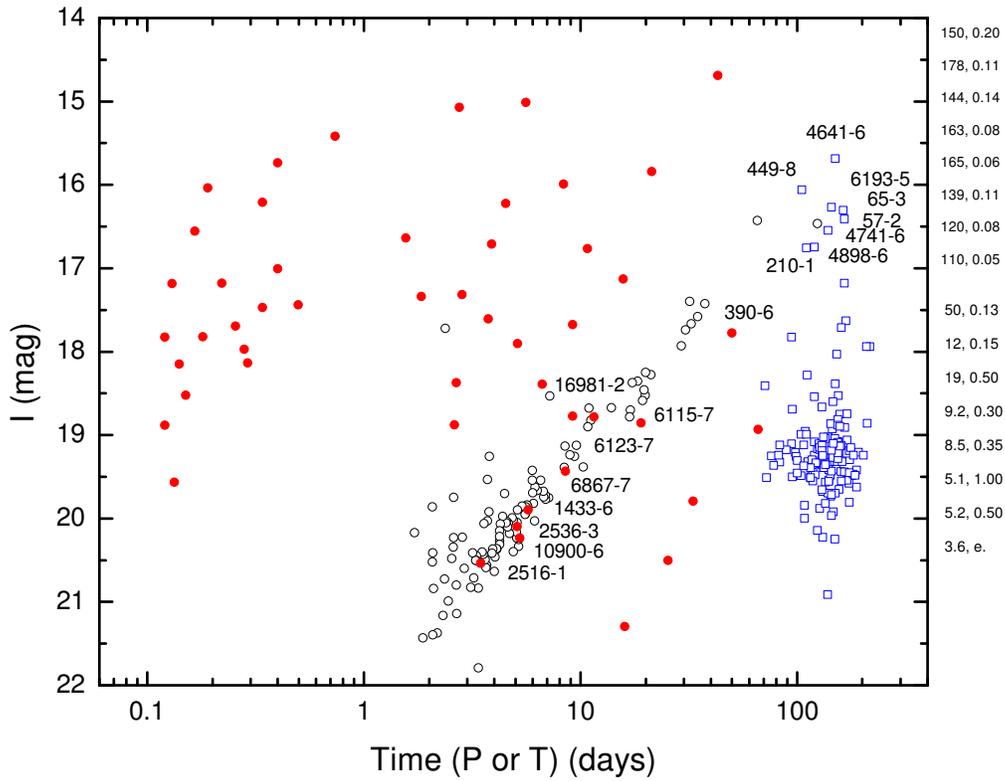}
      \caption{The "Cepheid-filtering" diagram. We include here true periods (dots)
      for periodic stars, and time scales of variability (squares) for probably periodic stars.
      The numbers at the right side correspond to the pairs [period (days), amplitude] for the Cepheid candidates.
         Cepheids found by P04 are shown as open circles.
         The new Cepheid candidates are labeled. Note the bright long-period candidates. They were not
         flagged as possible Cepheids in P04 due to their very small amplitudes.
              }
         \label{Fig*}
   \end{figure*}

  \begin{figure*}
   \centering
   \includegraphics[width=15cm]{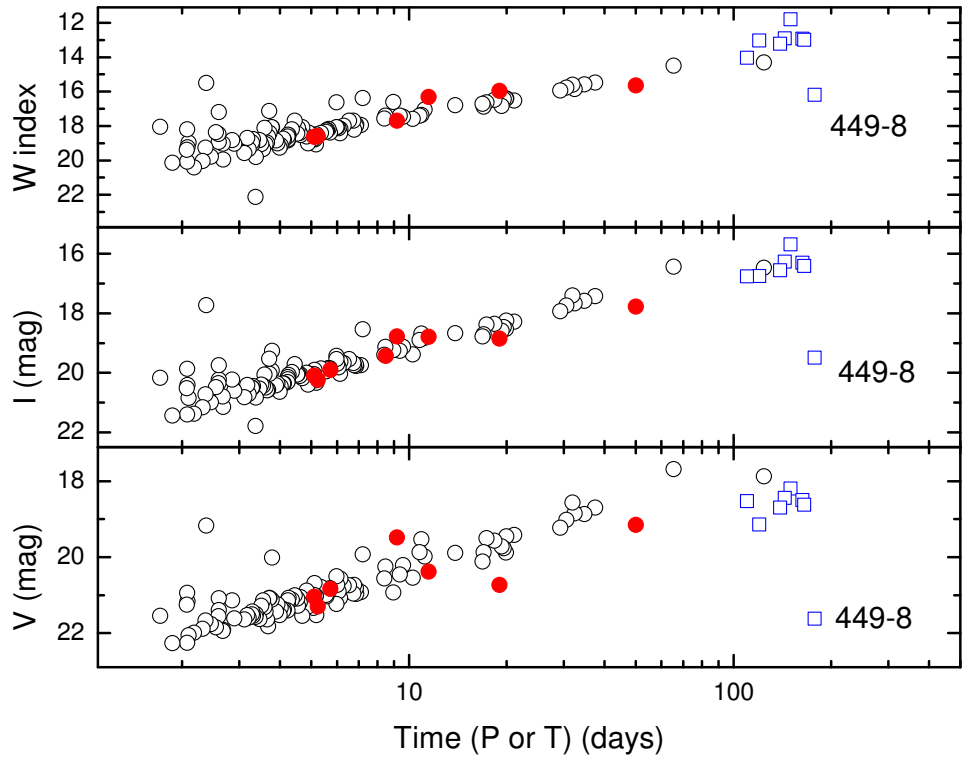}
      \caption{The $PL$ relationships (using the period, or time scale of variability) for
       the Cepheids of Pietrzynski et al. (2004, open circles),
       our periodic Cepheid candidates (filled circles)
       and our long-period, small-amplitude Cepheid candidates (open squares).  These diagrams
       show that, except for $449-8$, the long-period Cepheid candidates closely
       follow the $PL$ relationships in the I and reddening-free W bands, but
       are underluminous in the $V$ band.
      }
         \label{Fig}
   \end{figure*}

   \begin{figure*}
   \centering
   \includegraphics[width=15cm]{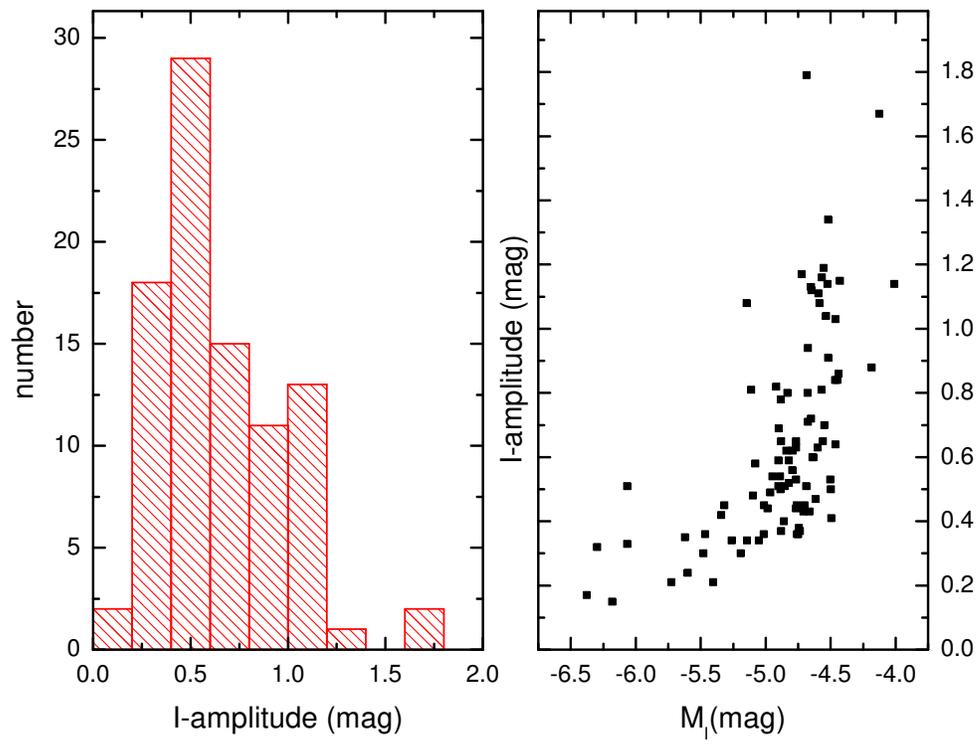}
      \caption{The left panel shows the histogram of the amplitudes for the variables in the bottom-right rectangle in Figure 3.
      The right panel shows the amplitude as a function of the absolute I-band magnitude.
              }
         \label{Fig}
   \end{figure*}

 \begin{figure*}
   \centering
   \includegraphics[width=15cm]{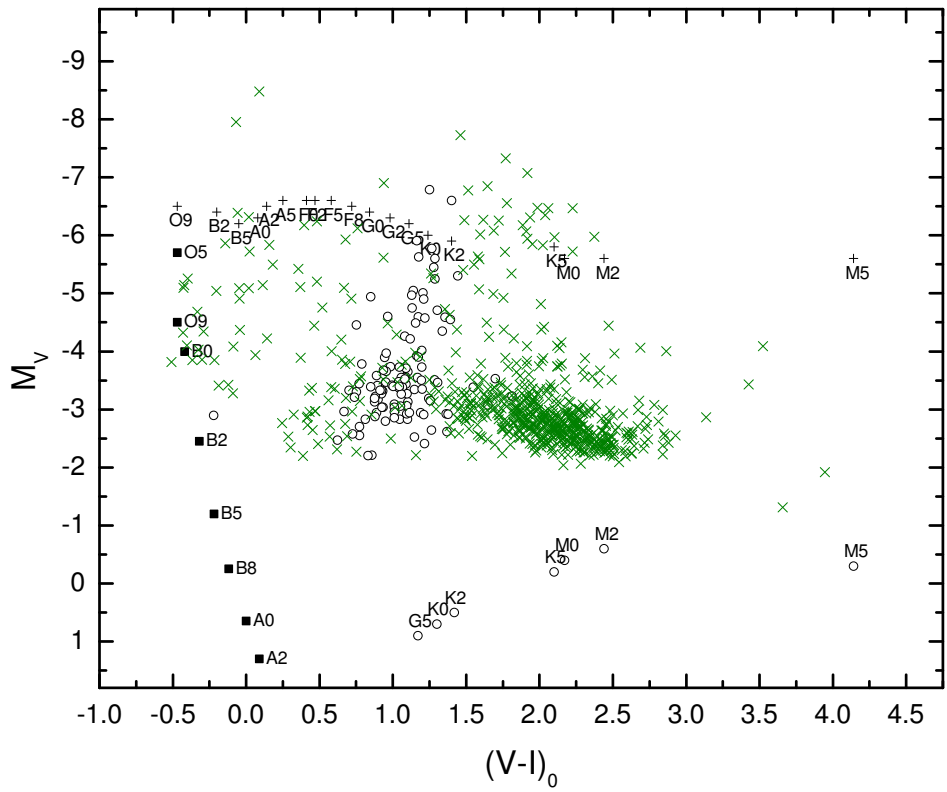}
      \caption{Same as Fig.\,3, but for "non-periodic" variable stars (crosses).
              }
         \label{Fig}
   \end{figure*}

    \begin{figure*}
   \centering
   \includegraphics[width=15cm]{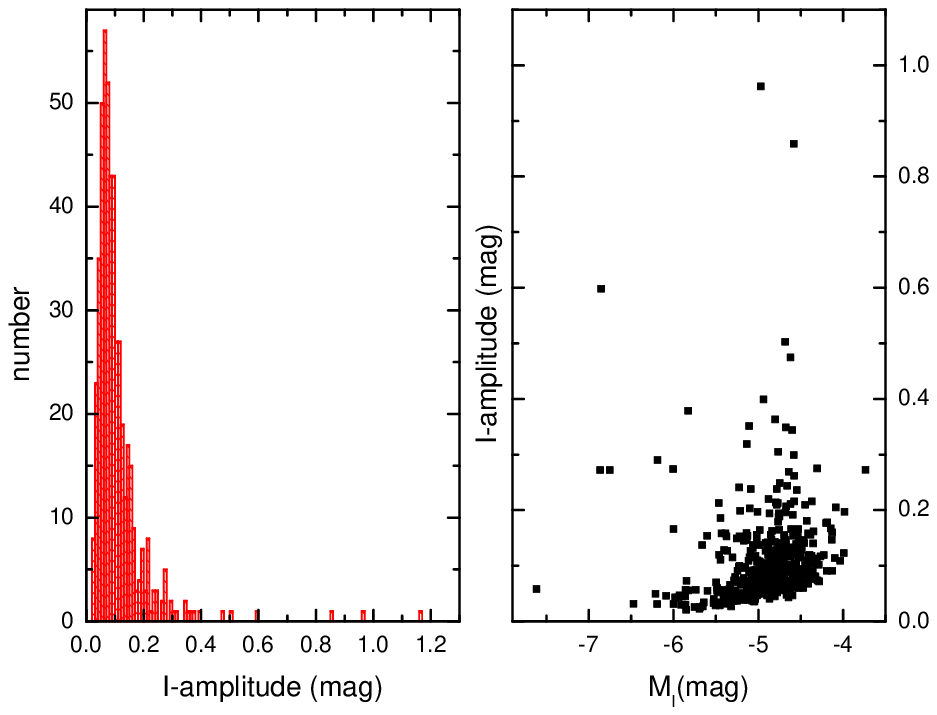}
      \caption{The left panel shows the histogram of the amplitudes for
      the red non-periodic variables in Figure 7
      (those with $M_{V} > -4.1$ and $(V-I)_{0} > 1.3$).
      The right panel shows the amplitudes plotted against the absolute I-band magnitudes
      for these stars.
              }
         \label{Fig}
   \end{figure*}

\section{Conclusions}

In this paper we have continued our search for extragalactic variable stars
making use of the image databases provided by the Araucaria Project
(for variable stars detected in NGC\,300 in the framework of this project, see Mennickent et al.\,2004; for details on
the Araucaria Project, see for instance Pietrzynski et al., 2002, and more recently Gieren et al., 2005).
With this study, we have significantly contributed to the
census and
understanding of the population of bright variable stars towards the galaxy NGC\,6822.
We have presented data for more than 1200 variable stars, far exceeding the number
of previously known objects of this kind  toward this galaxy.  Unfortunately,
for many objects
we cannot unambiguously determine their membership to NGC 6822, since
contamination by foreground Milky Way stars is a strong problem in such surveys
in Local Group galaxies. Our present survey is not an exception, especially due to the
low galactic latitude of NGC\,6822 ($b^{II}$ = -18$\degr$). According to the
Bahcall \& Soneira galaxy model for star densities,
we should expect  16.2 foreground stars
per square arcminute with magnitudes in the range of 13 $ < V < $ 23 towards NGC\,6822 (Ratnatunga \& Bahcall 1985).
This yields approximately 18\% of contamination for the total number of
stars detected in our survey. If we translate this figure to the detected variable stars,
 we would expect that 219 stars of our  1215 objects
could be galactic foreground stars. True membership to NGC\,6822
can only be established with spectroscopic follow-up surveys and, in some cases,
with multicolor photometry (Massey 1998). In any case, our work has been exploratory;
the detection of variable star candidates constitutes the first step towards a complete
understanding of the variable star population in this galaxy.
Our search was limited to $M_{V} < -2$, excluding pulsators like
RR-Lyrae and $\delta$ Scuti stars, but we should be able to detect $\beta$ Cephei,
Mira and semiregular variables, as well as RV Tau stars. Many of our targets could finally
be classified as these kind of variables once follow-up observations become available.
Further spectroscopic work
and  continuous photometric monitoring are therefore encouraged to understand the nature of these variables,
and to compare their properties with those of the corresponding populations in other galaxies.

An interesting byproduct of our investigation has been the discovery of a small population
of stars which could be strongly reddened, small-amplitude Cepheids of extremely long periods.
They seem to follow the Cepheid $PL$ relationship, but turn to be underluminous in the $V$ band.
We suspect that these 7 objects might be indeed very young Cepheids which are still embedded
in dusty envelopes. Further
work will be necessary to corroborate this hypothesis.

\begin{acknowledgements}
 We acknowledge the referee, Dr. Raffaele Gratton, for useful
comments on the first version of this manuscript.
R.E.M. acknowledges support by Grant \emph{Fondecyt}, project number
1030707. WG, GP and REM acknowledge financial support for this work from the Chilean Center for Astrophysics
\emph{FONDAP}, project number 15010003.
This work is based on data obtained with the Warsaw
telescope on Las Campanas. We gratefully acknowledge help of some of the members of the OGLE team
in the acqisition of these data.
\end{acknowledgements}

\end{document}